\documentclass[aps,super,twocolumn,nofootinbib,amssymb,amsmath,footinbib]{revtex4}
\usepackage{graphicx,}
\usepackage{dcolumn}% Align table columns on decimal point
\usepackage{bm}% bold math
\usepackage{natbib}

\newcommand{\be}{\begin{equation}}
\newcommand{\ee}{\end{equation}}
\newcommand{\bea}{\begin{eqnarray*}}
\newcommand{\eea}{\end{eqnarray*}}
\newcommand{\beal}{\begin{eqnarray}}
\newcommand{\eeal}{\end{eqnarray}}
\newcommand{\lx}{\lambda_x}
\newcommand{\ax}{\alpha_x}
\newcommand{\ay}{\alpha_y}

\begin{document}

\title{Timing and dynamics of single cell gene expression in the arabinose utilization system}

\author{Judith A. Megerle$^{\dagger,\star,1}$}
\author{Georg Fritz$^{\dagger,2}$}
\author{Ulrich Gerland$^{2}$}
\author{Kirsten Jung$^3$}
\author{Joachim O. R\"adler$^1$}

\affiliation{$^ \dagger $These authors contributed equally to this work.}
\affiliation{$^1$Department f\"ur Physik und CeNS, LMU M\"unchen, Germany}
\affiliation{$^2$Institut f\"ur Theoretische Physik, Universit\"at zu K\"oln, Germany\\ Present adress: Arnold Sommerfeld Center for Theoretical Physics und CeNS, LMU M\"unchen, Germany}
\affiliation{$^3$Department Biologie I, Bereich Mikrobiologie, LMU M\"unchen, Germany}
\affiliation{$^ \star $Corresponding author. Electronic adress: {\tt judith.megerle@lmu.de}}

% 200 words max Abstract
\begin{abstract}
The arabinose utilization system of {\it E. coli}
displays a stochastic ``all or nothing'' response at intermediate
levels of arabinose, where the population divides into a fraction
catabolizing the sugar at a high rate (ON state) and a fraction
not utilizing arabinose (OFF state). Here we study this decision
process in individual cells, focusing on the dynamics of the
transition from the OFF to the ON state. Using quantitative
time-lapse microscopy, we determine the time delay between inducer
addition and fluorescence onset of a GFP reporter. Through
independent characterization of the GFP maturation process, we can
separate the lag time caused by the reporter from the intrinsic
activation time of the arabinose system. The resulting
distribution of intrinsic time delays scales inversely with the
external arabinose concentration, and is compatible with a simple
stochastic model for arabinose uptake. Our findings support the
idea that the heterogeneous timing of gene induction is causally
related to a broad distribution of uptake proteins at the time of
sugar addition.

\end{abstract}

\maketitle

\section*{Introduction}

Bacteria have sophisticated signal transduction and gene
regulatory networks for rapid adaptation to environmental changes.
In recent years it became increasingly recognized, that the
dynamical response of these biochemical reaction networks is
subject to significant stochastic fluctuations \cite{Kaern2005},
which can lead to heterogeneous behavior across cellular
populations. Examples include the transient differentiation of
{\it B. subtilis} in its late exponential phase
\cite{Suel_Nature_06,Leisner_MolMicrobiol_07}, bacterial
persistence in {\it E. coli} \cite{Balaban2004}, and the mating
pheromone response pathway in yeast \cite{Paliwal_Nature_07}.  In
many of these systems, positive feedback plays a fundamental role,
since it gives rise to bistability and thereby causes two clearly
distinct gene expression states \cite{Isaacs_PNAS_03}. It has been
demonstrated that biochemical noise induces stochastic transitions
between the two ``stable'' states, and it was suggested that the
resulting population heterogeneity provides selective advantages
for colony growth in fluctuating environments
\cite{Thattai_Genetics_04,Kussell_Science_05}.

\begin{figure}
\centerline{\includegraphics[width=8.5cm]{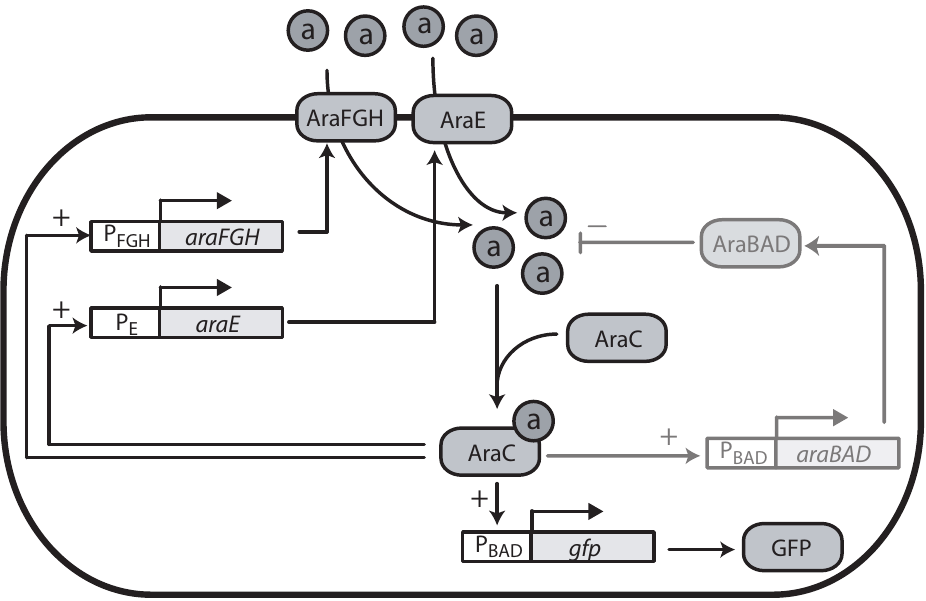}}
\caption{\label{FIGmodel}Regulatory network of
the native arabinose utilization system
\cite{Schleif_TrendsGenet_00}, including the {\it gfp}-reporter module
used in this study. The system consists of genes for arabinose uptake ({\it
araE, araFGH}), genes for arabinose metabolism ({\it araBAD}), and the regulator AraC. High amounts of intracellular arabinose activate AraC, which stimulates expression from the promoters $P_{BAD}$, $P_{E}$ and $P_{FGH}$.  In the absence of arabinose, AraC represses expression from $P_{BAD}$ (not depicted). Note that we also omitted the less pronounced negative autoregulation of AraC in the absence and presence of arabinose \cite{Johnson_JBacteriol_95}, since this feedback mainly seems to provide a constant transcription factor abundance \cite{Miyada_PNAS_84}. As indicated in light grey, in the mutant used in this study the chromosomal {\it araBAD} operon is deleted, and hence the additional  negative feedback on internal arabinose is avoided. As a reporter for the expression of the ara system  we used a plasmid-borne {\it gfp} variant under the control of the $P_{BAD}$ promoter, see Materials and Methods for details.}
\end{figure}

A prototypic class of positive feedback systems are the inducible sugar utilization systems, in which bistability is caused by the autocatalytic positive feedback of the sugar on its own uptake proteins. These systems allow bacteria to grow on less favorable carbon sources than glucose: For instance, in a medium where lactose is the only energy source, {\it E. coli}'s lactose utilization ({\it lac}) system either imports and catabolizes lactose at a high rate (ON state), or it does not use lactose at all (OFF state) \cite{Novick_PNAS_57}.  This bistable behavior has drastic effects on the behavior at the population level.  When a high amount of external lactose was added to a previously uninduced culture, all cells in the population switched from the OFF to the ON state. However, at lower sugar concentrations only a fraction of cells switched to the ON state while others remained in the OFF state \cite{Novick_PNAS_57, Ozbudak_Nature_04}.

Here, we are interested in the dynamics of such a switching process on the single cell level. We study these dynamics in the context of the arabinose utilization ({\it ara}) system of {\it E. coli} \cite{Schleif_TrendsGenet_00}, another well-characterized bistable system, see Fig.~\ref{FIGmodel}. In this case, arabinose is imported by the high-affinity low-capacity transporter AraFGH and the low-affinity high-capacity transporter AraE. If internal arabinose exceeds a threshold concentration, it activates AraC, which in turn promotes expression of {\it araFGH}, {\it araE} and the genes for arabinose catabolism {\it araBAD}. Siegele and Hu \cite{Siegele_PNAS_97} analyzed population distributions at intermediate sugar levels, and revealed that the {\it ara} system displays an all-or-nothing expression pattern similar to the {\it lac} system. They conjectured that in uninduced cells the stochastic background expression of the {\it ara} regulon leads to a wide distribution of {\it ara} uptake proteins. Addition of arabinose would then lead to different rates of arabinose accumulation, causing heterogeneous timing of gene induction within the population. At a given time there would be a fraction of induced and a fraction of uninduced cells, and the depletion of arabinose by the metabolism of the induced cells could explain the fixation of the all-or-nothing response. This conjecture is consistent with a computational study of autocatalytic expression systems \cite{Carrier_JTheorBiol_99} and experiments which placed {\it araE} under the control of a constitutive promoter, finding homogeneous gene expression in the population \cite{Khlebnikov_JBacteriol_00,Khlebnikov_Microbiology_01,Morgan-Kiss_PNAS_02}.
The dynamics of switching processes has also been studied using flow cytometry techniques, which yield a time series of population distributions of gene expression levels \cite{Mettetal_PNAS_06, Longo_Hasty_06}.

In this study we take a different experimental approach: Rather than recording population distributions, we use quantitative time-lapse fluorescence microscopy to follow the expression dynamics of the switching process in many cells, individually.
In a physics analogy, this is akin to following the trajectories of many particles, instead of recording their spatial density distribution at different time points. Clearly, the distributions can be obtained from the trajectories, but not vice versa, i.e. the trajectories contain more information.
In the present case, this additional information is particularly useful to disentangle different variables that affect the response of individual cells: The observed time-dependent fluorescence level is the final output of a series of biochemical processes, which can be grouped into two connected subsystems, an ``uptake module'' and a ``reporter module''. Both modules experience noise, which, to a first approximation, can be subsumed into a single parameter for each module. As we will see, one can extract these two parameters for each cell by fitting an appropriate model to the ``fluorescence trajectory'' of the cell. As a result, we can directly obtain the separate distributions for these two parameters, and even measure their correlations. Note that this analysis would not have been possible based on population distributions of gene expression levels.

Using this approach, we address the question raised by Siegele and
Hu \cite{Siegele_PNAS_97}, i.e. is the all-or-nothing response of
the {\it ara} system associated with heterogeneous timing of gene
induction, and, if so, is the heterogeneous timing causally
related to a wide distribution of {\it ara} uptake proteins? At 
subsaturating sugar levels, we observe a significant delay between
addition of inducer and increase of fluorescence, which is indeed
broadly distributed. To clarify the origin of this delay and its
broad distribution, it is necessary to separate the
intrinsic lag of the GFP expression dynamics from the time-lag
inherent to the stochastic arabinose uptake. To this end, we
leverage our microfluidic setup to separately measure the
distribution of GFP maturation times across an {\it E. coli}
population. We also record the cell-to-cell variation of the
growth rates. Using a simple quantitative model for the expression
dynamics, we then extract the intrinsic timing statistics for gene
induction. We find that this distribution is well described by an
analytical delay time distribution derived from a stochastic model
for the uptake module. Our results support the conclusion that the
heterogeneous timing is indeed due to a wide distribution of {\em ara}
uptake proteins across the population.

%%%%%%%%%%%%%%%%%%%%%%%%%%%%%%%%%%%%%%%%%%%%%%%%
\section*{Materials and Methods}
%%%%%%%%%%%%%%%%%%%%%%%%%%%%%%%%%%%%%%%%%%%%%%%%

\footnotesize

\paragraph*{\footnotesize Bacterial strain and plasmid.}
{\it E. coli} strain LMG194 [$F^{-}$ {\it lacX74 galE galK thi
rpsL phoA (PvuII) ara714} {\it leu::Tn10}] \cite{Guzman_JBacteriol_95}
was transformed with plasmid pBAD24-GFP (this work) using a
standard method as described elsewhere \cite{Promega1994}. The gene
{\it gfpmut3} \cite{Cormack_Gene_96} encoding the Green Fluorescent Protein GFPmut3 was
amplified by PCR with primers GFP-KpnI sense
(5'-TACCATGGTACCAAGTAAAGGAGAAGAACTTTTC-3') and \- GFP-HindIII
antisense (5'-CATAG\-T\-A\-AG\-C\-TTTTATTTGTAT\-A\-G\-T\-T\-CATCCATGCC-3') using
plasmid pJBA29 \cite{Andersen_ApplEnvMicrobiol_98} as a template.
The DNA-fragment was cut with restriction endonucleases KpnI and
HindIII, and was then ligated into similar treated vector pBAD24
\cite{Guzman_JBacteriol_95}, resulting in plasmid pBAD24-GFP. The
correct insertion of the fragment was verified by restriction
analysis as well as by DNA sequence analysis.

\paragraph*{\footnotesize Growth conditions.}
Cells were grown in LB medium \cite{Sambrock_89} or M63
minimal medium \cite{Guzman_JBacteriol_95} containing 0.2 \% (w/v)
glycerol as C-source. When indicated, 0.01\%, 0.02\%, 0.05\% or 0.2\% (w/v)
arabinose was added to induce GFP expression. Bacteria were
inoculated from single colonies grown on LB agar plates and grown
overnight ($37^\circ$C, shaking at 300rpm) in M63 medium.
Overnight cultures were diluted 1:50 into fresh M63 medium and
cultured for 2\,h. Bacteria were subsequently diluted in prewarmed
medium to an appropriate density and were then applied to one
channel of a Poly-L-Lysine coated microfluidic chamber
($\mu$-Slide VI, ibidi, Martinsried, Germany). The slide was then
incubated at $37^\circ$C for several minutes. By softly flushing
the channel with prewarmed medium supplemented with the desired
arabinose concentration, gene expression was induced and the sample
was rinsed at the same time. After the preparation procedure the
vast majority of the bacteria adhered with their long axis
parallel to the surface.

\paragraph*{\footnotesize Time-lapse microscopy.}
Time-lapse experiments were performed on a fully automated
inverted microscope (Axiovert 200M, Zeiss, Oberkochen, Germany)
equipped with a motorized stage (Prior Scientific, Cambridge, UK).
All devices were controlled by Andor IQ software (Andor, Belfast,
Northern Ireland). Fluorescence illumination was provided by an
X-cite120 light source (EXFO, Quebec, Canada). An appropriate
filter set (excitation: 470/40; beamsplitter 495; emission:
525/50; filter set Nr 38, Zeiss, Oberkochen, Germany) was used.
Bright field and fluorescence images of several fields in one
sample were acquired every 5\,min with a highly sensitive EMCCD
camera (iXon DV885, Andor, Belfast, Northern Ireland) through an
oil-immersion 100x plan-neofluar objective with NA 1.3 (Zeiss,
Oberkochen, Germany), with acquisition times of 0.1\,s to 0.2\,s.
To further prevent photobleaching and photodamage all light
sources were shuttered between exposures and an orange filter was
used in the bright field light path. The temperature in the sample
environment was maintained at $37^\circ$C using a custom built
heating box. Focalcheck fluorescence microspheres (Invitrogen,
Karlsruhe, Germany) were used to correct for output variations of
the lamp.

\paragraph*{\footnotesize Data Analysis.}
ImageJ \cite{Abramhoff_BiophotInt_04} and Igor Pro 4.0
(WaveMetrics, Lake Oswego, OR) were used for data analysis. Cell
outlines were created by thresholding the bright-field images. Total
fluorescence was measured as the sum over all pixel values within
the outline in the corresponding background corrected fluorescence
image. Time traces were assembled by tracking the cells manually.
As photobleaching was found to be negligible for the given
experimental system, fluorescence traces were fitted without
further processing.

\paragraph*{\footnotesize Measurement of the GFP maturation time distribution {\it in vivo}.} The maturation time in single cells was determined using an approach similar to the one established in Ref.~\cite{Gordon_NatMeth_07}: Translation was blocked by the addition of 200\,${\mu}g/ml$ chloramphenicol, 30 min after the induction of {\it gfp}-expression with 0.2\% arabinose. Fluorescence images were acquired every 3 to 5\,min before and after inhibition. As this
measurement was more sensitive the illumination was reduced and
the EM Gain of the camera was used. Photobleaching could thus
again be neglected. Cellular fluorescence was determined by
summing all pixel values above the background level for each
bacterium. This method is qualitatively equal to the use of cell
outlines as described above, but can only be applied if the range
of fluorescence values is limited and bacteria do not grow
strongly. The resulting maturation time courses were fitted by an
exponential function.

\normalsize

%%%%%%%%%%%%%%%%%%%%%%%%%%%%%%%%%%%%%%%%%%%%%%%%
\section*{Results}
%%%%%%%%%%%%%%%%%%%%%%%%%%%%%%%%%%%%%%%%%%%%%%%%

\subsection*{Single Cell Induction Kinetics}

To study the induction kinetics of the {\it ara} system, we use an
{\it E. coli} strain where both {\it araBAD} and {\it araC} are
deleted \cite{Guzman_JBacteriol_95}. It is transformed with the
reporter plasmid pBAD24-GFP, containing the {\it araC} gene
and the rapidly maturing GFP variant {\it gfpmut3} \cite{Cormack_Gene_96} which is under the
control of the $P_{BAD}$ promoter, see Materials and Methods. The {\it araC} gene is supplied on the plasmid to guarantee full functionality of the DNA loop required for repression of $P_{BAD}$ in the absence of arabinose \cite{Schleif_TrendsGenet_00} and to provide the proper stoichiometry of transcription factors and $P_{BAD}$ promoters. The chromosomal deletion of {\it
araBAD} avoids the negative feedback of the internal arabinose
catabolism. This feedback complicates the system, but is
irrelevant for our questions, which focus on the kinetics of the
induction when arabinose first becomes available externally. The
gene regulatory circuit of our system is illustrated in
Fig.~\ref{FIGmodel}.

\begin{figure}
\centerline{\includegraphics[width=7.1cm]{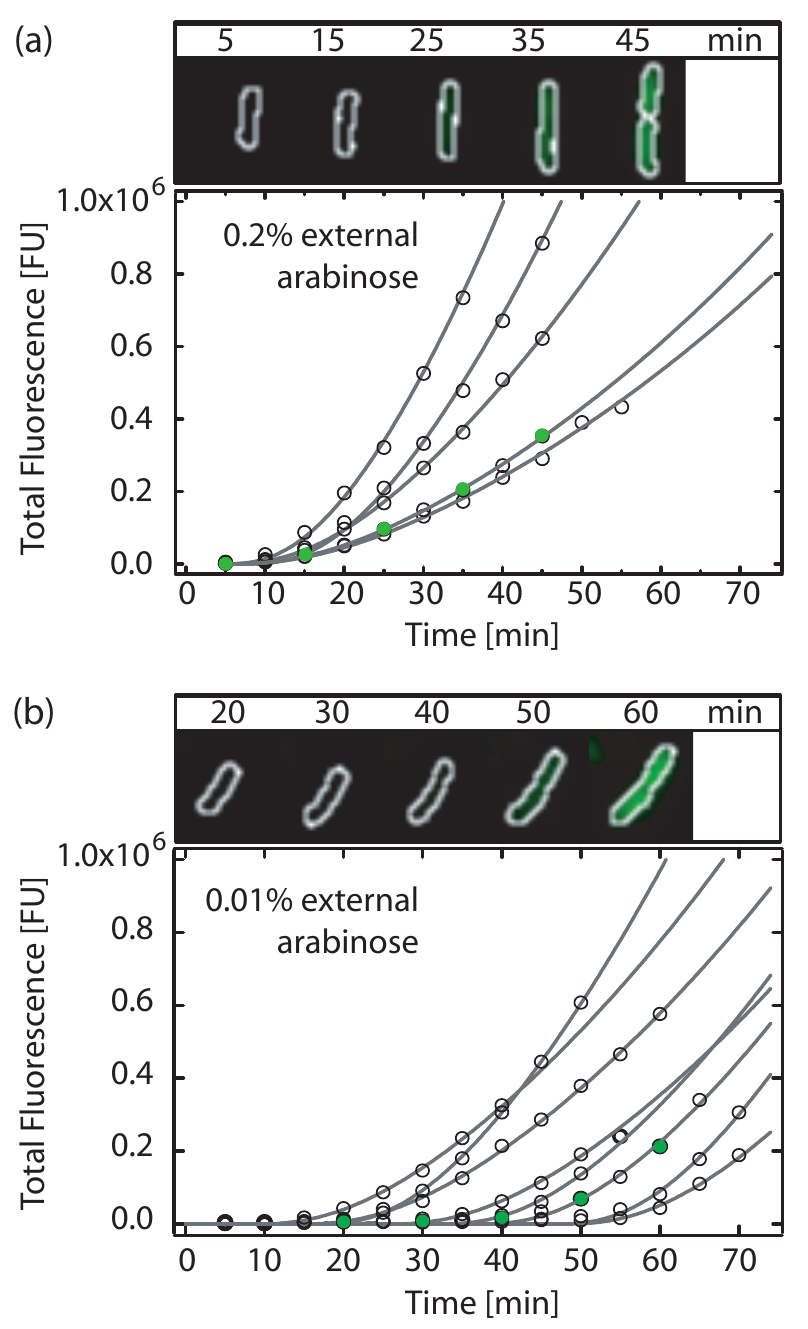}}
\caption{\label{FIGtimeseries}Examples of single cell induction kinetics of the arabinose utilization network. Cells were induced at t=0\,min with 0.2\% arabinose (a) and 0.01\% arabinose (b)  ({\it empty circles}). The traces were analyzed up to the first cell
division, which results in different numbers of data points in the
traces.
Fits of the deterministic gene expression function in
Eq.~\ref{EQNgfp} to the data are shown as solid lines.
The image panels in (a)
and (b) correspond to the fluorescence traces marked with green circles, respectively. The total  fluorescence was determined within the white outlines created via thresholding of the respective bright field images and is given in fluorescence units (FU).}
\end{figure}

To perform the time-lapse fluorescence microscopy, we introduce
the bacteria into a microfluidic chamber, where they attach to the
Poly-L-Lysine coated chamber wall. The microfluidic chamber
provides homogeneous external conditions for the bacteria and can
be used to rapidly exchange the medium. At t=0 min, we induce the
bacteria with 0.2\% (13.3\,mM), 0.05\% (3.33\,mM), 0.02\%
(1.33\,mM) or 0.01\% (0.66\,mM) arabinose, and then record the
time-evolution of GFP fluorescence in single cells. Representative
fluorescence ``trajectories'' for the highest (0.2\%) and the lowest (0.01\%) arabinose concentration are shown in Fig.~\ref{FIGtimeseries}~(a) and (b), respectively.

For all arabinose concentrations, the individual time-traces of
each cell appear rather smooth and deterministic, whereas there is
a significant variation in the response from cell to cell. We also
observe a time lag between the addition of arabinose and
the onset of fluorescence. With decreasing arabinose 
concentration, the typical lag time becomes longer, and its cell to 
cell variation becomes more pronounced. 
Below, we will devise a rigorous way to quantify this delay. 
Here, we only apply a simple thresholding procedure to extract 
an apparent lag time. Using an intensity threshold of 
$2.5\cdot10^4$ fluorescence units (FU), we determine an 
apparent lag time of $16\pm2.5$\,min at 0.2\% arabinose 
and a more substantial delay of $34\pm10$\,min at 0.01\% 
arabinose. In the latter case approximately 10\% of the 
bacteria do not show any fluorescence within our time 
window of 70\,min.

With the sudden increase of the external arabinose concentration
at t=0 min, a cascade of biochemical processes is triggered,
culminating in the fluorescent output signal measured in our
experiment. In order to narrow down the origin of the stochasticity
in the apparent lag time, we need to analyze the individual steps
in this cascade. For this analysis, it is useful to separate the
system into two distinct modules, an uptake module and a GFP
expression module, as depicted in Fig.~\ref{FIGSchemas}~(a). The
uptake module not only comprises arabinose import (represented
here by an effective uptake protein ``Upt'' that subsumes
transport by AraE and AraFGH), but also includes the positive
feedback of arabinose on the uptake protein. The expression module
turns ON the production of the output signal, when internal
arabinose reaches a threshold level \cite{Schleif_JMB_69}. The
delay time $\tau_D$ that is required to reach this threshold is
solely determined by the uptake module. However, GFP fluorescence
does not follow promoter activation instantaneously. Instead, the
processes of transcription, translation and GFP maturation
depicted in Fig.~\ref{FIGSchemas}~(b) also generate a dynamical
delay and thereby contribute to the apparent delay estimated
above. To quantitatively estimate the intrinsic delay $\tau_D$ and
its statistics, we now scrutinize the expression module in detail.

%%%%%%%%%%%%%%%%%%%%%%%%%%%%%%%%%%%%%%%%%%%%%%%%%%
\subsection*{Quantitative characterization of the expression module}

\begin{figure}
\centerline{\includegraphics[width=8.5cm]{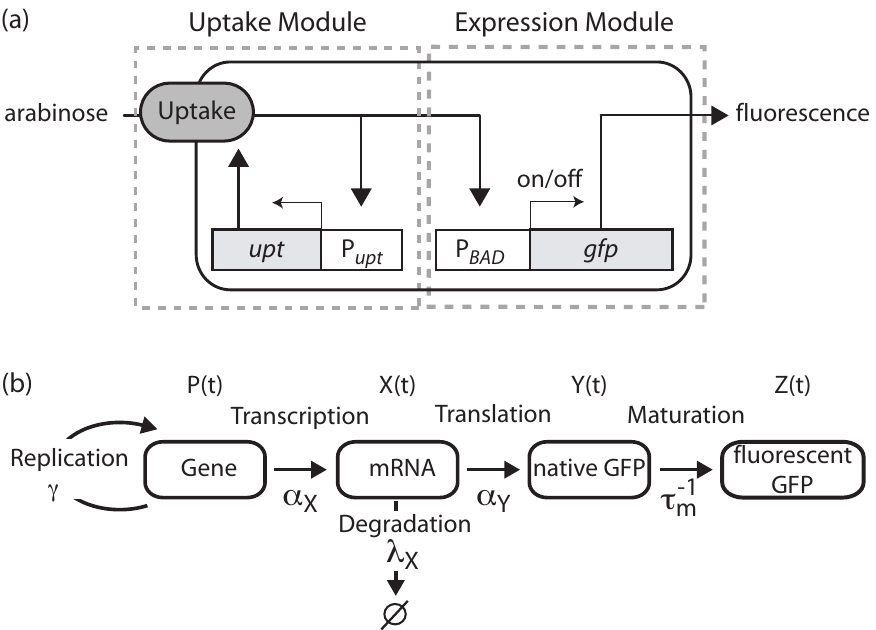}}
\caption{\label{FIGSchemas}
(a) The arabinose utilization system can be dissected into an arabinose uptake module ({\it left}) and a gene expression module ({\it right}). The gene expression module is turned ON, if the internal arabinose level exceeds the threshold required for activation of the $P_{BAD}$ promoter. The stochastic model for the uptake module comprises arabinose import by a heuristic uptake protein and the positive feedback of arabinose on the synthesis of the uptake protein, see Appendix~B for all details. The model for the expression module encompasses the processes depicted in (b) and describes the accumulation of total fluorescent GFP per cell, see Appendix~A for the deterministic rate equations.}
\end{figure}

\begin{figure}
\centerline{\includegraphics[width=6.5cm]{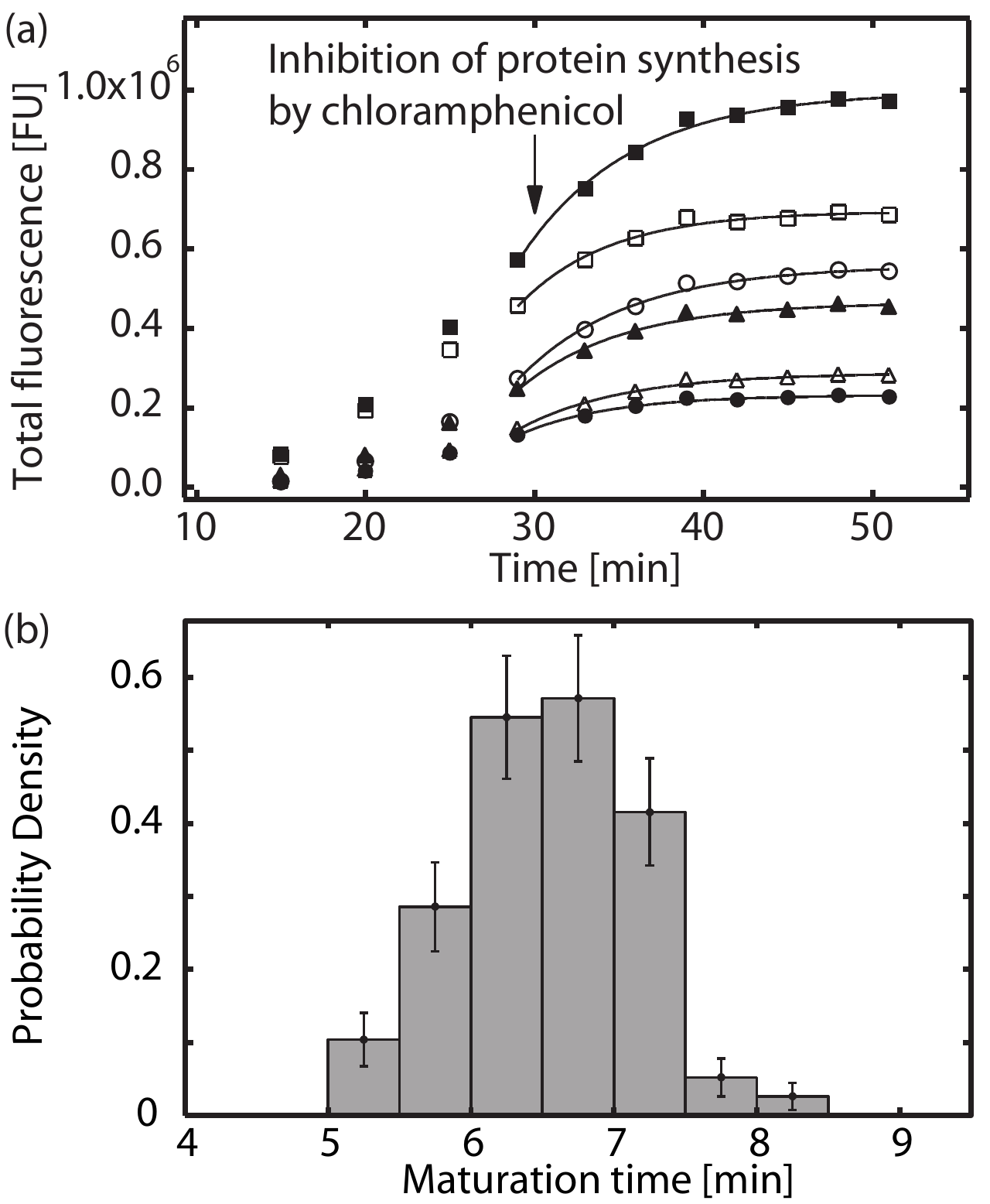}}
\caption{\label{FIGmaturation} 
GFP maturation kinetics in single cells. In (a) GFP expression was induced with 0.2\% arabinose at t=0\,min and protein synthesis was inhibited by addition of 200 $\mu g/m l$ chloramphenicol at t=30\,min, as indicated by the arrow. Exponential fits to the fluorescent timeseries ({\it solid lines}) yield the maturation-time distribution in (b). The statistics was obtained from 77 cells.}
\end{figure}

\paragraph*{GFP maturation time.}
A significant portion of the dynamic delay of the expression module is incurred by GFP maturation, the process whereby the folded protein becomes fluorescent. The rate-limiting reaction is an oxidation with a time constant of several minutes up to several hours \cite{Tsien_AnnuRevBiochem_98}, depending on the variant of the protein and possibly on the organism. However, for our present purpose, we not only need the average time constant, but also need to know whether there is a large cell-to-cell variation associated with the maturation process. With our microfluidic setup, we can directly probe this cell-to-cell variation experimentally, under the same conditions as in the induction experiments. First, we induce bacteria with 0.2\% arabinose and then inhibit protein synthesis {\it in situ} by flushing the channel with the antibiotic chloramphenicol. The resulting fluorescence trajectories cease to increase about 15 min after the addition of the antibiotic, see Fig.~\ref{FIGmaturation}~(a) for a few representative trajectories. Following the rationale established in Ref.~\cite{Gordon_NatMeth_07}, this behavior reflects the maturation dynamics of the remaining, non-fluorescent GFPs. The distribution of time-constants $\tau_m$ of GFP maturation shown in Fig.~\ref{FIGmaturation}~(b) was obtained from exponential fits to 77 single-cell timeseries ({\it solid lines} in Fig.~\ref{FIGmaturation}~(a)).
We find an average maturation time of $\tau_m = 6.5$\,min and a standard deviation of $0.6$\,min, i.e. a cell-to-cell variation of only about 10\%.

Our finding of a relatively small cell-to-cell variation suggests that the maturation process is largely independent of the internal state of the cell in {\it E. coli}. This appears plausible, given that the oxidation reaction does not depend on intracellular components \cite{Tsien_AnnuRevBiochem_98}. For comparison, measurements of the maturation times of YFP and CFP in yeast \cite{Gordon_NatMeth_07} found considerably longer maturation times of $\sim 40$~minutes, but only a slightly larger relative cell-to-cell variation ($15-20\,\%$). Moreover, from {\it in vitro} measurements of various YFP variants, oxidation timescales as low as 2-8~minutes were determined \cite{Nagai_NatBiotechnol_02}, indicating that the rapid maturation time detected in our experiment is conceivable {\it in vivo}.

\paragraph*{Gene copy number.}
Since our GFP reporter is encoded on a plasmid, the average copy-number of the plasmid and its cell-to-cell variation are important properties of the expression module.
The plasmid pBAD24 has an average copy number comparable to pUC \cite{Cronan_Plasmid_06}, which is present in about 55 copies per cell \cite{Lin-Chao_MolMicrobiol_92}. Assuming plasmid production and dilution with constant rates, we expect Poissonian fluctuations on the order of $\sqrt{55} \approx 7.5 $ plasmids (13\,\%). In similar plasmids, ColE1 and R1, negative feedback is known to reduce the copy-number variations below the Poisson limit \cite{Paulsson2001}. This may also apply to pBAD24, which would make the variation even less significant.
We expect that the plasmid copy number grows proportional to the volume of the cell, such that the {\em concentration} of plasmids remains constant. Hence, we will assume that the rate $\gamma$ of gene replication in Fig.~\ref{FIGSchemas}~(b) equals the rate of volume expansion of the cells.

\paragraph*{Cell growth.}
As the above discussion of the gene copy number shows, the distribution of growth rates is another characteristic affecting the quantitative properties of the expression module.
We analyzed the growth of individual cells in the microfluidic channel by recording the time-evolution of their area detected under the microscope. Since the rod-shaped {\it E. coli} cells grow mainly along their principal axis (cf. image panels in Fig.~\ref{FIGtimeseries}), the growth rate of the cell area is a proxy for the growth rate by cell volume. From exponential fits \cite{Cooper_BMCBiology_06} to 84 timeseries of the cell area we found a distribution of time constants for cell growth with an average of 50~min and a standard deviation of 6~min. Hence, the cell-to-cell variations of the growth rate is also relatively small. This result indicates that the microchemical conditions in our channel are sufficiently constant to guarantee a reproducible growth state of the cells. We also found that the doubling time was independent of the arabinose concentration, consistent with the fact that in this strain arabinose cannot be catabolized and used as an energy source.

\paragraph*{mRNA half-life and protein expression rate.}
Finally, the dynamics of the expression module is dependent on the rate constants for {\it gfp} expression and mRNA degradation.
Average mRNA half-lifes were determined for most of {\it E. coli}'s genes \cite{Bernstein_PNAS_2002} and are typically in the range 3 to 8\,min. The work of Smolke et al. \cite{Smolke_ApplEnvMicrobiol_00} indicates that the population-averaged half-life of {\it gfp} mRNA is in the same range; for our analysis below we will assume an average half-life of 6\,min. In contrast, there is currently no report on the cell-to-cell variation of {\it gfp} mRNA half-lifes. We expect that such a variation would mainly be produced by cell-to-cell variations of RNase abundance and other components required for transcript turnover. These components have been shown to vary with the growth rate \cite{Nilsson_Nature_84}. Since the growth rate varies only by $\sim$10\% from cell to cell in our experiment (see above), we estimate the relative cell-to-cell variations of mRNA half-life to be similar. This may be an overestimate, since the degradation machinery negatively autoregulates its own expression \cite{Jain_GenesDev_95}, a mechanism known to reduce gene expression noise \cite{Becskei_Nature_00}.

The protein expression rate has been quantified experimentally at the single-cell level for the  $P_R$ promoter of phage $\lambda$, and substantial cell-to-cell variations on the order of 35\% were determined \cite{Rosenfeld_Science_05}. These large relative differences likely stem from cell-to-cell variations in global cellular components such as RNA polymerases or ribosomes. We expect similar variations for GFP expression from the $P_{BAD}$ promoter.

\begin{figure}
\centerline{\includegraphics[width=6.5cm]{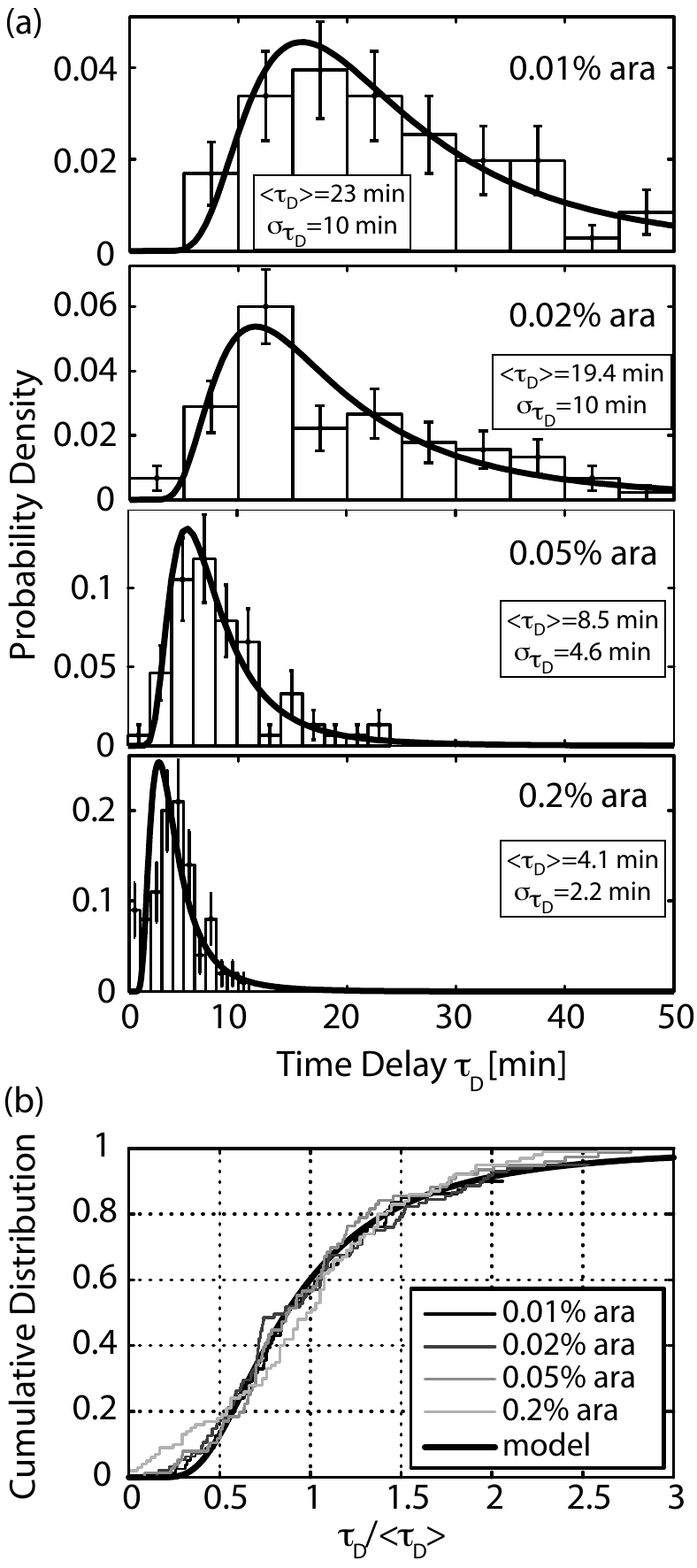}}
\caption{\label{FIGdistribution}
(a) Histograms of the time delay $\tau_D$ for varying external arabinose concentrations, as determined from the fits of Eq.~\ref{EQNgfp} to the fluorescence timeseries. The mean $\langle\tau_D\rangle$ as well as the standard deviation $\sigma_{\tau_D}$ gradually decrease for increasing arabinose levels. Note that for 0.01\% arabinose approximately 10\% and for 0.02\% arabinose approximately 5\% of the cells did not turn on gene expression within our experimental time window. Therefore the extraced means and standard deviations (see {\it insets}) constitute slight underestimates in these cases. The solid lines are fits of the analytical delay time distributions (Eq.~\ref{EQNdelay_time_distr}) to the data, for details see text. The statistics was obtained from 71 cells at 0.01\%, 90 cells at 0.02\%, 76 cells at 0.05\%, and 101 cells at 0.2\% arabinose. (b) Cumulative distributions of the delay times rescaled to their mean values $\langle\tau_D\rangle$. A two-sample KS-test indicates that all rescaled distributions are likely to be drawn from the same underlying probability distribution. The p-values of the individual pairs are 0.25 for (0.01\% and 0.02\% arabinose), 0.87 for (0.01\% and 0.05\% arabinose), 0.54 for (0.01\% and 0.2\% arabinose), 0.90 for (0.02\% and 0.05\% arabinose), 0.08 for (0.02\% and 0.2\% arabinose), and 0.67 for (0.05\% and 0.2\% arabinose).  
The analytical prediction is shown for $\mu = 3.8$, $b=30$, and $\tau_0 = 2100$\,min ({\it bold line}).}
\end{figure}

%%%%%%%%%%%%%%%%%%%%%%%%%%%%%%%%%%%%%%%%%%%%%%%%%%
\subsection*{Distribution of GFP expression rate and intrinsic delay time}

Given the above characterization of the expression module, we can now construct a simple quantitative model for its dynamic response, and then use this model to extract the intrinsic delay $\tau_D$. The smooth shape of the timeseries in Fig.~\ref{FIGtimeseries} suggests that the dynamics of individual cells follows a rather deterministic fate, while the differences between the cells stem from cell-to-cell variation of the reaction rates. Therefore we use a deterministic rate equation model to describe the expression dynamics within a single cell, but allow for cell-to-cell variation in the model parameters. This model follows the reaction scheme depicted in Fig.~\ref{FIGSchemas}~(b):
Transcription of {\it gfp} mRNA from the promoter $P_{BAD}$ is turned ON at $t=\tau_D$ and then remains constant at rate $\alpha_x$. However, the number of plasmids (and hence gene copies) increases with rate $\gamma$, which equals the cell doubling rate, so that the plasmid copy number $P$ remains stable in the bacterial population. We denote the mRNA degradation rate by $\lambda_x$, and the translation and maturation rates of GFP by $\alpha_y$ and $\tau_m^{-1}$, respectively (see Appendix~A for details).

Within this model, the time-evolution of the total number of fluorescent GFP molecules in a cell, $Z(\tau)$, is described by the expression
\begin{eqnarray}
Z(\tau)   &  =  & \alpha_p\left( \frac{(\gamma + \lambda_x)e^{-   \tau/\tau_m}}{(\gamma + \tau_m^{-1})( \lambda_x - \tau_m^{-1})} \right .\nonumber\\
& & \left. + \frac{  \tau_m^{-1} e^{- \lambda_x\, \tau}}{\lambda_x
(\tau_m^{-1} - \lambda_x)} + \frac{ \tau_m^{-1}  e^{\gamma\,
\tau}}{\gamma (\gamma + \tau_m^{-1})}  \right) - Z_0\, ,
\label{EQNgfp}
\end{eqnarray}
where  $\tau = t - \tau_D$ is the time after transcription is switched ON, $\alpha_p \equiv P \alpha_x \alpha_y / (\gamma + \lambda_x)$ is a lumped constant giving the protein synthesis rate in fluorescence units per minute [FU/min], and $Z_0$ is a constant determined by the initial conditions. Here, the first two terms in parentheses describe transients associated with the equilibration of the GFP maturation process and the mRNA degradation reaction, respectively, i.e. their contributions decay exponentially with time constants $\tau_m$ and $\lambda_x^{-1}$. In the long-time limit the last, exponentially increasing term is dominant. It reflects the constant protein production from an exponentially growing number of plasmids, and describes the long-time behavior of the {\it total} fluorescence per cell. However, since we study the dynamics of gene expression during the first cell cycle after induction, all terms, including the transients, are relevant.

From the previous section, we conclude that the parameter $\alpha_p$, comprising the plasmid copy number and the protein expression rate, captures most of the cell-to-cell variation within the expression module. To fit the model in Eq.~\ref{EQNgfp} to the single-cell induction kinetics, we therefore fixed the remaining parameters to their population-averaged values.
Hence, in the optimization procedure of the fit, we only allow the adjustment of $\alpha_p$ and the uptake-induced delay $\tau_D$, which we seeked to extract. Note that this choice fixes all relevant timescales governing the dynamics in Eq.~\ref{EQNgfp} and the free parameters only impose shifts in the onset ($\tau_D$) and in the absolute magnitude ($\alpha_p$) of {\it gfp}-expression.

We fitted the timeseries of cells induced with various levels of arabinose (0.2\%, 0.05\%, 0.02\%, and 0.01\%). A few representative fitted curves for the highest and lowest concentration are plotted in Fig.~\ref{FIGtimeseries} as solid lines. The resulting histograms for the delay time are shown in Fig.~\ref{FIGdistribution}~(a). For the lowest arabinose level (0.01\%, {\it upper panel}) we find that the delay times are distributed between 5 and 50\,min with a mean and standard deviation of $\langle \tau_D\rangle =23$\,min and $\sigma_{\tau_D} = 10$\,min, respectively. With increasing arabinose concentration both the mean and the standard deviation of the delay time distribution decrease gradually, until at the highest arabinose level (0.2\%, {\it lower panel}) a distribution with $\langle \tau_D \rangle =4.1$\, min and $\sigma_{\tau_D} =  2.2$\,min is reached.

\begin{figure}
\centerline{\includegraphics[width=7.6cm]{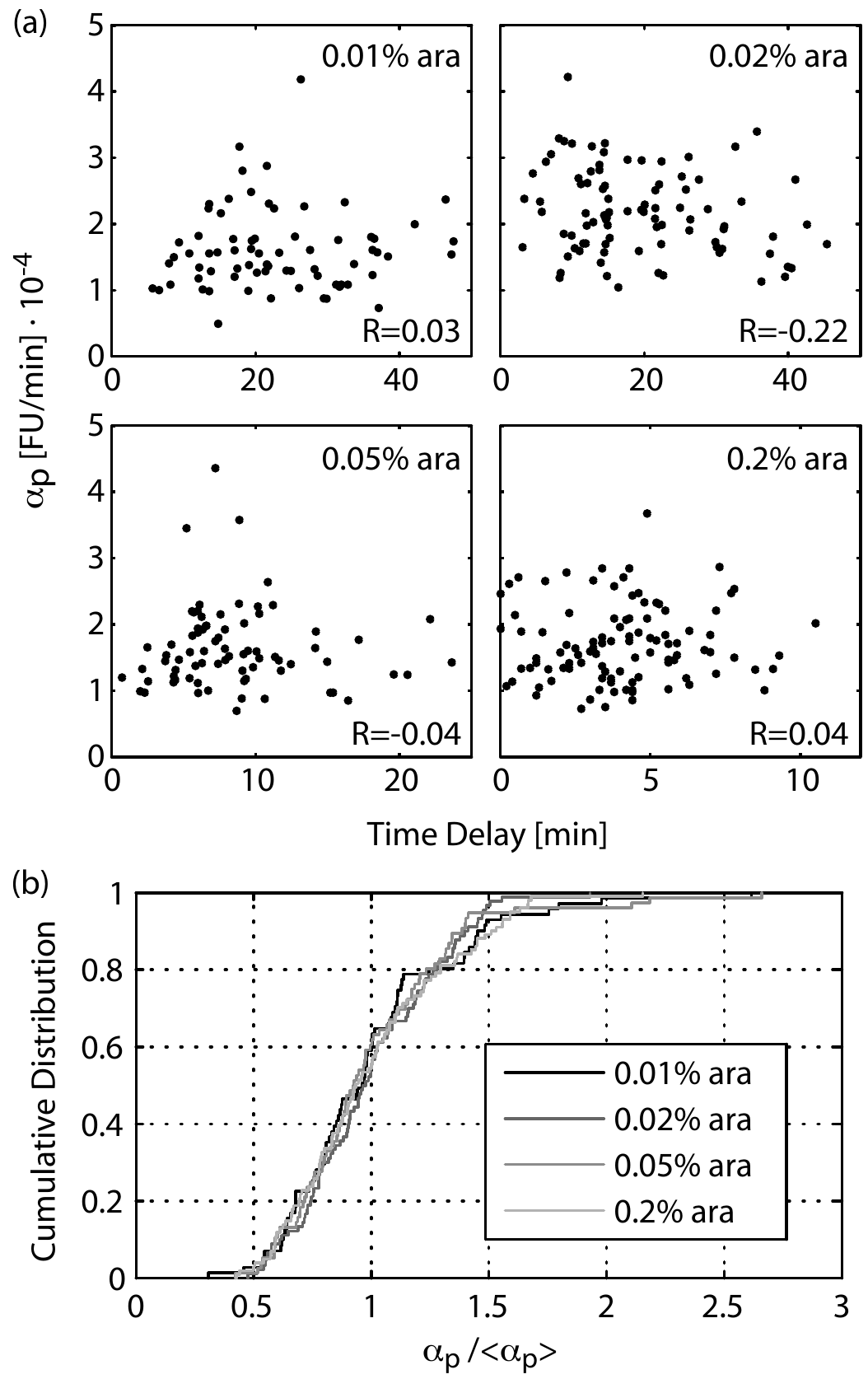}}
\caption{\label{FIGscatterplots}
(a) Correlations between delay time and protein synthesis rate $\alpha_p$. The scatter plots display small correlation coefficients R, and the respective p-values for observing these correlations by random chance are 0.68 for 0.01\%, 0.035 for 0.02\%, 0.73 for 0.05 \%, and 0.72 for 0.2\% arabinose. (b) The cumulative distributions of the protein synthesis rate $\alpha_p$ were rescaled to their mean values $\langle \alpha_p\rangle$ to exclude sample-to-sample variations of the mean. Importantly, we found no correlations between $\langle \alpha_p\rangle $ and the inducing arabinose concentration. A two-sample KS-test shows that all rescaled distributions of $\alpha_p$ are compatible with each other. The significance levels of the pairwise KS-tests varied between 0.57 and 0.97.
}
\end{figure}

To test whether there is a relationship between the delay time and the protein synthesis rate, we calculated their cross-correlation coefficients for all inducing arabinose levels, see Fig.~\ref{FIGscatterplots}~(a). Only in the case of 0.02\% arabinose a slight anticorrelation was detected, whereas for all other concentrations the correlation coefficient is close to zero [p-values for finding the observed correlation coefficients by chance in an uncorrelated sample: 0.68 for 0.01\% ara, 0.03 for 0.02\% ara, 0.73 for 0.05\% ara, and 0.72 for 0.2\% ara]. We also find that the distribution of {\it gfp}-expression rates itself does not vary systematically with the inducing arabinose concentration, and all distributions fall on top of each other when rescaled by their mean values, see Fig.~\ref{FIGscatterplots}~(b) [pairwise Kolmogorov-Smirnov (KS) tests yield significance levels between 0.57 and 0.97 for the null hypothesis that the data sets are drawn from the same underlying distribution]. In summary, the low correlations between $\tau_D$ and $\alpha_P$ on the one hand and the independence of $\alpha_P$ on the inducing arabinose level on the other hand, suggest that the uptake and the expression module are indeed functionally separate. Note that our experimental approach with time-lapse fluorescence microscopy was crucial for these results, which would have been impossible to obtain with flow cytometry.

\subsection*{Stochastic model for the uptake module}

Next, we want to assess whether the extracted delay time distributions of Fig.~\ref{FIGdistribution}~(b) may be causally linked to a broad variation in the number of uptake proteins. We approach this question with the help of a simple stochastic model for the ``uptake module'' depicted in Fig.~\ref{FIGSchemas}~(a).
The model is useful in three respects:
(i) It serves us to illustrate the mechanism whereby stochastic expression of the uptake protein genes can produce a broad distribution of delay times. We will see that according to this mechanism, the delay time distributions for different inducer concentrations should be related by simple linear rescaling of the time axis. Thus, we will test for this signature of the mechanism in our experimental data.
(ii) Since most model parameters are strongly constrained by literature values, we can test whether an interpretation of our data based on the stochastic model is consistent with these constraints.
(iii) Independent of the precise choice of parameter values, which affect the average delay time and its standard deviation, the model predicts a certain shape for the delay time distribution. We will test whether this shape is compatible with our data.

There are two distinct transport systems for arabinose uptake, AraE and AraFGH. However, the two systems are coupled, and it was found that arabinose uptake can effectively be described as a single Michaelis-Menten process \cite{Daruwalla_BiochemJ_81}. In the sketch of Fig.~\ref{FIGSchemas}~(a), this combined transport system is represented by a single gene ``{\it upt}''.
In addition to the transport, the uptake module of Fig.~\ref{FIGSchemas}~(a) comprises the activation of AraC by internal arabinose, the subsequent stimulation of transcription by the activated complex, and the translation into functional uptake protein. Within our stochastic model for the uptake module, we describe and simulate all of these processes in standard ways, see Appendix~B for details.

\begin{figure}
\centerline{\includegraphics[width=7.6cm]{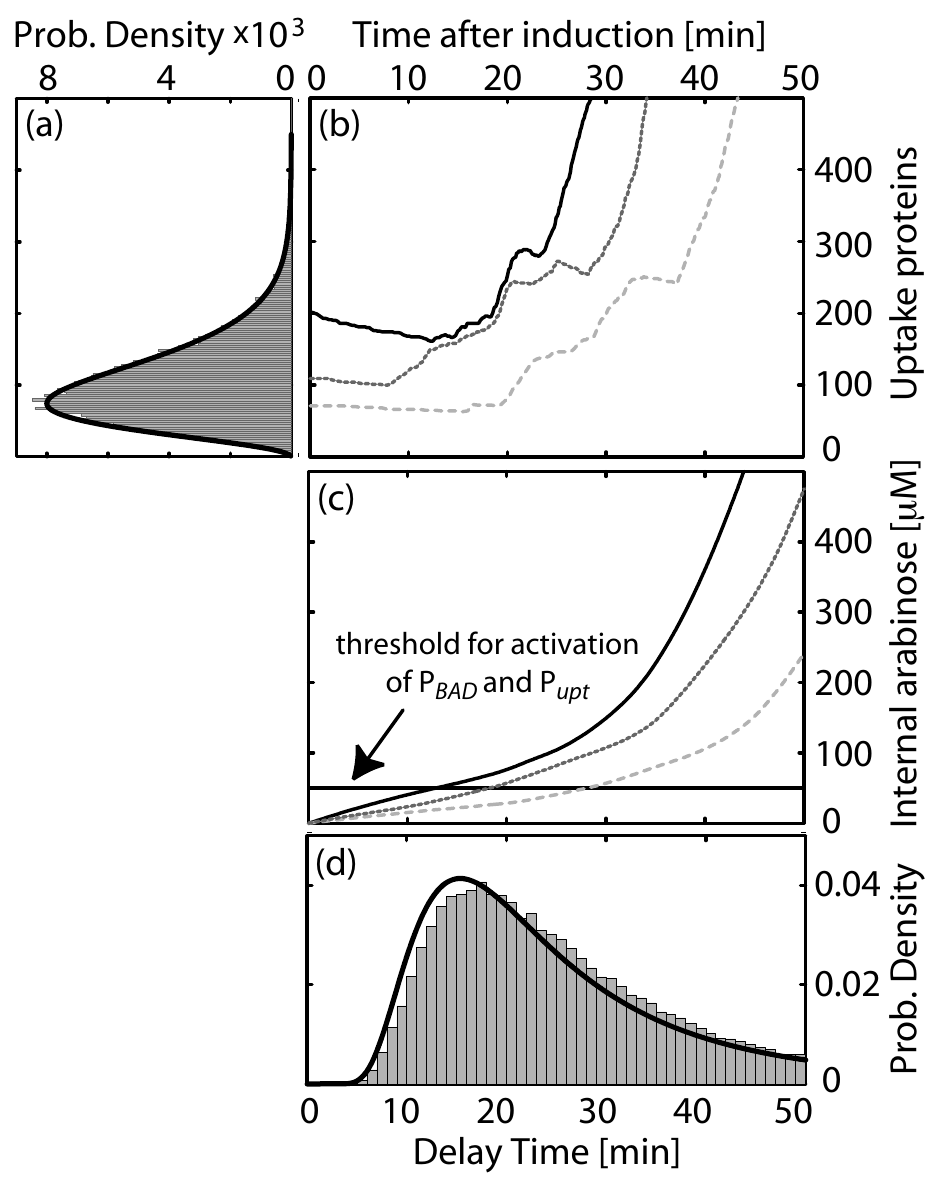}}
\caption{\label{FIGtimeseries_arabinose}Illustration of the stochastic arabinose uptake mechanism at 0.01\% external arabinose (simulation).
The three representative time-courses of arabinose uptake proteins in (b) and internal arabinose in (c) illustrate that the rate of arabinose uptake is proportional to the amount of uptake protein present at a given time. Once the internal threshold for activation of the promoters $P_{BAD}$ and $P_{upt}$ is reached, the positive feedback gets activated and is visible as the kinks in (b) and (c). The delay time distribution in (d) ({\it grey bars}) is obtained by measuring the time to reach this threshold. If the uptake proteins decay much slower than the typical time required to reach the threshold (adiabatic limit), the delay time distribution in (d) can be related to the steady state distribution of uptake proteins at zero arabinose in (a). The analytical predictions in (a) and (d) ({\it bold lines}) are shown for $\mu=3.8$, $b=30$, and $\tau_0=2100$\,min (for details see text).}
\end{figure}

Figs.~\ref{FIGtimeseries_arabinose}~(b) and (c) show the simulated time-evolution of the level of uptake proteins and the level of internal arabinose upon induction with 0.01\% external arabinose for a few representative simulation runs. These trajectories illustrate the mechanism leading to a broad distribution of delay times within our model: Internal arabinose initially accumulates approximately linearly in time, and the accumulation accelerates only after reaching the effective arabinose threshold of $a_0 \approx 50\, \mu M$ for activation of the {\it araBAD} and {\it upt} promoters, which is indicated by the black horizontal line in Fig.~\ref{FIGtimeseries_arabinose}~(c). The time delay, $\tau_D$, caused by the uptake module is the time required for the internal arabinose concentration to reach this threshold level. The rate of arabinose import, given by the slope in Fig.~\ref{FIGtimeseries_arabinose}~(c), is proportional to the number of uptake proteins $n$ in Fig.~\ref{FIGtimeseries_arabinose}~(b). If arabinose import is fast compared to the timescale of changes in the protein abundance, the delay time is given by the simple relation $\tau_D = a_0/(v_0 n)$, where the arabinose uptake rate per uptake protein, $v_0$, depends on the external arabinose concentration. Thus, the distribution of uptake proteins in Fig.~\ref{FIGtimeseries_arabinose}~(a) directly determines the distribution of import rates, which in turn are inversely proportional to the delay times, resulting in the distribution of delay times shown in Fig.~\ref{FIGtimeseries_arabinose}~(d).

A simple prediction of this mechanism is that an increase of the uptake velocity $v_0$ will reduce all delay times within a distribution of cells by the same factor.
In other words, the delay time distributions for different arabinose levels (and hence different $v_0$) should fall on top of each other upon simple linear rescaling of the time axis (and restoring normalization). In Fig.~\ref{FIGdistribution}~(b) we test this prediction on our experimental time delay distributions. We find that after rescaling to the same mean value, the cumulative distributions are congruent with each other. 
This agreement is also quantitatively supported by pairwise KS-tests, which test whether the samples are likely to be drawn from the same underlying distribution (the legend to Fig.~\ref{FIGdistribution} shows the respective significance levels). 
Note that the linear scaling of the time axis with $1/v_0$ does not imply linear scaling with the arabinose level, since $v_0$ depends nonlinearly on the external arabinose level, see also further below.

In order to relate the experimentally observed shape of the distribution to the prediction of the stochastic model, we will now derive an analytical expression for the delay time distribution. To this end, we first consider only intrinsic noise and study the effect of extrinsic noise below. Before the addition of the inducer arabinose, expression of the uptake proteins is a completely random, unregulated process. Following the work of Berg \cite{Berg_PNAS_78} and under the assumptions stated in Appendix~B, we find a steady-state distribution $P(n)$ for the number of uptake proteins $n$ of the form
\begin{equation}
\label{EQNprotein_distribution}
P(n) \approx \left( \frac{1}{1+b}\right)^\mu\,\left( \frac{b}{1+b}\right)^n\,\left( \begin{array}{c} \mu+n-1\\n \end{array} \right)\,,
\end{equation}
which is sometimes referred to as a `negative binomial'.
Here, the ratio $b= \nu_p/\lambda_m$ of the translation rate and the mRNA degradation rate corresponds to the typical number of proteins produced from a single mRNA and is also known as the ``burst size'' \cite{Thattai_PNAS_01}. The ratio $\mu=\nu_m^0/\lambda_p$ of the basal transcription rate and the protein dilution rate can be interpreted as a dimensionless ``burst frequency'' (the number of bursts within the lifetime of a protein). Both parameters determine the mean $\langle n \rangle = \mu b$ and the variance $\delta n^2 = \langle n \rangle (1+b)$ of $P(n)$. Fig.~\ref{FIGtimeseries_arabinose}~(a) shows the steady-state distribution $P(n)$ obtained from our stochastic simulations of the uptake module ({\it grey histogram}) together with the analytical expression in Eq.~\ref{EQNprotein_distribution} for the same rate constants. The excellent agreement suggests that the assumptions leading to Eq.~\ref{EQNprotein_distribution} are all satisfied in the relevant parameter regime.

\begin{figure}
\centerline{\includegraphics[width=7cm]{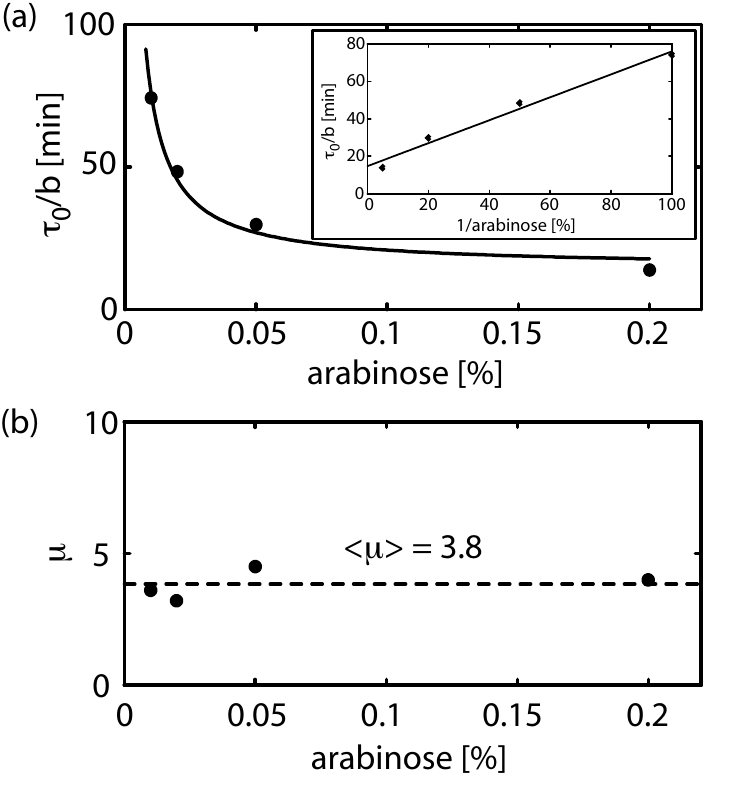}}
\caption{\label{FIGfitted_parameters}Estimated parameters as a function of external arabinose, as obtained from fits of the delay time distributions in Fig.~\ref{FIGdistribution}~(a). The timescale of arabinose accumulation $\tau_0/b$ in (a)  decreases monotonically with the inducing arabinose concentration, as expected from saturating Michaelis-Menten kinetics of the uptake proteins, cf. the Lineweaver-Burk plot ({\it inset}) for the scaling with the inverse arabinose concentration. In contrast, the burst frequency $\mu$ in (b) is constant for all arabinose levels. This is consistent with our central assumption that the underlying uptake protein distribution responsible for the heterogeneous timing is independent of the inducing arabinose concentration.}
\end{figure}

Next, we study the effect of extrinsic noise which leads to a variation of reaction parameters from cell to cell. An experimental characterization of extrinsic noise in {\it E. coli} \cite{Rosenfeld_Science_05} found a typical parameter variation of $\sim 20\%$. When we adopt this level of extrinsic noise for all parameters in our stochastic simulations, the resulting protein distribution has a significantly larger standard deviation than the distribution in the absence of extrinsic noise, while the mean remains almost unchanged, see {\it Suppl. Fig.~S1}. However, the protein distribution in the presence of extrinsic noise is still well fitted by Eq.~\ref{EQNprotein_distribution}, with an increased effective burst size and a reduced effective burst frequency. Keeping this in mind, the following results can be generalized to the realistic scenario where extrinsic fluctuations are present.

To obtain an approximation for the delay time distribution, we assume that arabinose uptake is rapid compared to the typical timescale of changes in the protein abundance. In this adiabatic limit, the delay time is inversely proportional to the current protein abundance in each cell, i.e. $\tau_D = \tau_0/ n$, where $\tau_0\equiv a_0/v_0$ is the time for a single uptake protein to accumulate arabinose to the threshold level $a_0$. With this relation, the steady-state uptake protein distribution (Eq.~\ref{EQNprotein_distribution}) leads to a delay time distribution of the form
\begin{equation}
\label{EQNdelay_time_distr}
Q(\tau_D) \approx \frac{\tau_0}{\tau_D^2}\,\left( \frac{1}{1+b}\right)^\mu\,\left( \frac{b}{1+b}\right)^{\tau_0/\tau_D} \frac{\Gamma (\tau_0/\tau_D+\mu)}{\Gamma (\tau_0/\tau_D + 1) \Gamma (\mu) }\,,
\end{equation}
where $\Gamma(x)$ is the Gamma function. In Fig.~\ref{FIGtimeseries_arabinose}~(d) we compare this analytical prediction ({\it black line}) to the stochastic simulation ({\it grey bars}). The small deviation stems from the fact that the number of uptake proteins is not constant over the period of the time delay\footnote{Indeed, if the protein dynamics is much faster than the characteristic time of arabinose uptake ($\lambda_p^{-1}\ll \tau_D$), every cell experiences simply the average abundance of uptake protein $\langle n \rangle$ and the delay time distribution approaches a sharply peaked function around $\tau_D = \tau_0\langle n \rangle^{-1}$ (data not shown). In our case, $\lambda_p^{-1}\approx 70$\,min is much larger than the average delay times, so that the assumption of a constant $n$ is sufficiently accurate.}. The mean and variance of the delay time distribution can be approximated by
\begin{eqnarray}
\langle \tau_D\rangle & \approx &  \frac{\tau_0}{\langle n \rangle}\,\left(1+\frac{\delta n^2}{\langle n \rangle^2}\right) \approx \frac{\tau_0}{\mu b}\,\left(1+\frac{1}{\mu}\right),\; \nonumber\\
\delta \tau_D^2 &\approx& \frac{\tau_0^2}{\langle n \rangle^2} \frac{\delta n^2}{\langle n \rangle^2}  \approx \left(\frac{\tau_0}{\mu b}\right)^2 \frac{1}{\mu} \,,
\label{EQNmean_var}
\end{eqnarray}
see Appendix~B.
From these expressions it is clear that the model has two key parameters, which together determine the mean and width of the delay time distribution: the time required to reach the internal arabinose threshold by a single protein burst, $\tau_0/b$, and the burst frequency $\mu$.

Now we test whether the shape of the delay time distribution predicted by the model is quantitatively consistent with our experimental distributions. To this end, we fit the model in Eq.~(\ref{EQNdelay_time_distr}) to the data in Fig.~\ref{FIGdistribution}~(a) by varying the two key parameters identified above. The resulting fits ({\it solid lines}) display good agreement with the experimental data, as indicated by one-sample KS-tests under the null hypothesis that the samples are drawn from the analytical distribution. The significance levels are 0.50, 0.47, 0.77, and 0.07 for 0.01\%, 0.02\%, 0.05\%, and 0.2\% arabinose, respectively. Only in the case of 0.2\% arabinose the test points to a significant difference between the theoretical and experimental distribution. However, for this concentration the estimated delay times are very short, such that the error of the estimation itself is likely to account for the deviations. Note that the two-parameter fit guarantees that the mean and standard deviation of the experimental and theoretical distribution will match. However, the fact that the {\em shape} of the distributions show excellent agreement is a nontrivial result, suggesting that the discussed delay mechanism can indeed explain our observations.

Finally, we address the consistency of the parameter values. Fig.~\ref{FIGfitted_parameters} shows the estimated parameters as a function of the external arabinose concentration: The timescale $\tau_0/b$ of arabinose accumulation in (a) decreases monotonically as a function of external arabinose and saturates for large sugar abundances, whereas the burst frequency $\mu$ in (b) is constant for all arabinose levels. This observation is consistent with the idea that the underlying protein distribution, characterized by $\mu$ and $b$, is independent of the externally provided sugar concentration, and that the differences in timing can be explained by shifts in the effective arabinose uptake velocity per uptake protein, $v_0$: By assuming simple Michaelis-Menten saturation kinetics for $v_0$, one expects that $\tau_0$ scales inversely with the external arabinose concentration $[a_{ex}]$, i.e. $\tau_0 =a_0/v_{max} \left(1+ K_m/[a_{ex}]\right)$, where $v_{max}$ denotes the maximal uptake velocity per uptake protein  and $K_m$ the Michaelis constant. This behavior is indeed found in Fig.~\ref{FIGfitted_parameters}~(a) ({\it inset}) and with the resulting values for $v_{max}$, $K_m$ and a typical value of $b=30$ for the burst factor \cite{Thattai_PNAS_01}, all parameters are compatible with the experimentally constrained ranges discussed in Appendix~B.

%%%%%%%%%%%%%%%%%%%%%%%%%%%%%%%%%%%%%%%%%%%%%%%%%
\section*{Discussion}
%%%%%%%%%%%%%%%%%%%%%%%%%%%%%%%%%%%%%%%%%%%%%%%%%

We studied the expression dynamics during induction of the bistable arabinose utilization system in single {\it E. coli} cells using quantitative time-lapse fluorescence microscopy. Upon addition of arabinose, we observed a characteristic time delay before the cells switched from a state of basal expression to a state of high expression of the {\it ara} regulon. This typical duration of the delay exhibited a systematic dependence on the externally supplied arabinose concentration:
At a saturating arabinose level, we found rapid induction within all cells of the culture, whereas with decreasing levels, we detected a significant broadening and shift of the delay time distribution function.
In order to characterize the cell-to-cell variability in the cellular response, we dissected the system into an uptake module with stochastic behavior, and an expression module which displays virtually deterministic behavior in individual cells.
We first studied the expression module, in particular by measuring the cell-to-cell distribution of the GFP maturation time. To the best of our knowledge, this constitutes the first measurement of a maturation time {\it distribution} in bacteria.
We then developed a hybrid deterministic/stochastic theoretical model to analyze our experimental data. The model is based on the assumption that the initial basal expression of the arabinose transporters determines the rate of arabinose uptake.
Adopting the approach of Berg, we find an analytic expression for the distribution of transporter proteins and the distribution of delay times. The theory consistently fits the shape of the experimental delay time distributions for various inducer concentrations.
Hence our data support a previous conjecture by Siegele and Hu \cite{Siegele_PNAS_97}, according to which the delay time distribution is causally linked to the distribution of uptake proteins in the absence of the inducer.
To corroborate our model even further it would be interesting to control the level of transporter proteins independently, e.g. by using an inducible promoter that is independent of arabinose. Also, it remains an open question how the two transport systems are coupled. It appears that the high-affinity low-capacity transporter {\it araFGH} and the low-affinity high-capacity transporter {\it araE} are orchestrated to respond like a single protein. A similar analysis to ours using knockout mutants in one of the two transport systems could shed light on this matter.

In general, we determined the dynamic response of bacteria to an external change of food conditions. Since such decisions are of vital importance to living systems, we can speculate about their impact on the fitness of a bacterial population. The observed heterogeneous timing in gene induction may simply be
a fortuitous consequence of the evolutionary process that shaped
the arabinose utilization system in {\it E. coli}. Alternatively,
it may be beneficial for a bacterial colony, if the individual
cells respond at different times when arabinose suddenly becomes
available in modest amounts. Note that in our experiments with the
{\it araBAD} deficient strain, even the lowest arabinose level, if
maintained over a long time, ultimately induces the {\it ara} system in
almost all cells. However, for a wild-type strain in an
environment where arabinose availability may fluctuate, temporal
disorder of gene induction could provide selective advantages for
the colony as a whole. For instance, it might be beneficial to
prevent costly synthesis of the arabinose system in all cells when
the sugar level is only moderate and may soon be depleted. Our
analysis indicates that the delay time distribution of the system
can be readily tuned over evolutionary timescales, by adjusting
the burst frequency and burst size of the uptake proteins. In the
future, it will be interesting to further explore the possible
connections between the system design in individual cells and the biological function at the population level.

\section*{Appendix}

\subsection*{Appendix A: Deterministic GFP expression model}
\footnotesize
To extract the intrinsic time delay $\tau_D$ from our single cell expression data, we employ a simple deterministic model that follows the scheme depicted in Fig.~\ref{FIGSchemas}~(b). We assume that the transcription rate from the promoter $P_{BAD}$ is zero until the internal arabinose threshold for activation of $P_{BAD}$ is reached at $t=\tau_D$. Then, the promoter activity jumps to its maximal value $\alpha_x$.
The corresponding rate-equations for the total
abundance of plasmids ($P$), {\it gfp} mRNA ($X $), immature GFP
protein ($Y$), and mature GFP protein ($Z$) per cell, are
\begin{eqnarray}
\partial_t  P& = & \gamma P \nonumber\\
\partial_t  X& = & \ax P  - \lx  X  \nonumber\\
\partial_t  Y& = & \ay X - \tau_m^{-1}  Y \nonumber\\
\partial_t  Z& = & \tau_m^{-1}  Y \;. \label{EQNmodel} \nonumber
\end{eqnarray}
with the cell-doubling rate $\gamma$ and the rate for transcription $\ax$, translation $\ay$, maturation $\tau_m^{-1}$, and mRNA degradation $\lx$. Note that the model does not include dilution due to cell growth, since we measured the total fluorescence per cell in our experiments. Therefore the number of plasmids (number of gene copies) increases exponentially in time, keeping the number of genes per volume constant.
Solving these equations for $Z(\tau)$ leads to Eq.~\ref{EQNgfp} in the main text.
\normalsize

\subsection*{Appendix B: Stochastic model for arabinose uptake}
\footnotesize
The arabinose uptake module, see Fig.~\ref{FIGSchemas}~(a), includes the processes for the uptake of arabinose as well as transcription, translation, and turnover of uptake proteins. In the following we describe the chemical reactions included in the stochastic simulations used to generate Fig.~\ref{FIGtimeseries_arabinose}~and Suppl.~Fig.~S1. We then derive an analytical approximation for the delay time distribution and discuss the experimental constraints on the model parameters.

{\it Arabinose uptake.} Comparison of arabinose uptake in wildtype strains with {\it araE} and {\it araFGH} deletion strains revealed that the two transporters do not operate independently \cite{Daruwalla_BiochemJ_81}. Instead, arabinose transport was best described by a single Michaelis-Menten function. Our model reflects this behavior of the wildtype strain through the use of a single ``effective'' uptake protein (referred to as Upt) for arabinose import,
\begin{eqnarray}\label{EQstoch_mod}
 a_{ex} +\mathrm{Upt} & \stackrel{K_{m}}{\longleftrightarrow} &  a_{ex}\cdot \mathrm{Upt} \nonumber \\
 a_{ex}\cdot \mathrm{Upt} &  \stackrel{v_{max}}{\longrightarrow}  & a + \mathrm{Upt }\nonumber \, .
 \end{eqnarray}
The uptake protein binds external arabinose $a_{ex}$ with dissociation constant $K_m$ and, once bound, translocates it to the cytoplasm at rate $v_{max}$. The effective uptake velocity per uptake protein is hence $v_0=v_{max} [a_{ex}]/(K_m+[a_{ex}])$. Cytoplasmic arabinose is denoted by $a$.

{\it Transcriptional regulation.} The $P_{BAD}$ promoter in the {\it ara}-regulon is one of the best characterized bacterial promoters: In the presence of internal arabinose, AraC stimulates transcription from $P_{BAD}$, while AraC represses transcription by formation of a DNA loop in the absence of arabinose \cite{Schleif_TrendsGenet_00}. When exceeding an arabinose threshold of $a_0\approx50\,\mu$M, the promoter activity of $P_{BAD}$ increases cubically with the internal arabinose concentration \cite{Schleif_JMB_69}. In contrast to the detailed studies on $P_{BAD}$, less is known about the promoter activity function of the promoters $P_E$ and $P_{FGH}$, which regulate expression of the transport proteins. Both promoters are also induced by internal arabinose, but lack an upstream AraC-binding site (required for DNA looping) and are not repressed in the absence of arabinose. Consequently, their basal expression level is higher than for $P_{BAD}$ and the fold-change is reduced from $\sim$400 for $P_{BAD}$ to $\sim$150 for $P_E$ and $P_{FGH}$ \cite{Kolodrubetz_JMolBiol_81}. However, the detailed promoter activity as a function of internal arabinose is not known for these promoters. Apart from the lack of the AraC binding site required for DNA looping, the promoters $P_E$ and $P_{FGH}$ display a high similarity to $P_{BAD}$ \cite{Johnson_JBacteriol_95}. Therefore we model transcriptional regulation of the uptake proteins by introducing a heuristic promoter $P_{upt}$, that has the same characteristics as $P_{BAD}$, but lacks the repression in the absence of arabinose. To reproduce the cubic increase of the promoter activity function of $P_{BAD}$ we allow three arabinose molecules to bind AraC with dissociation constant $K_C$. This activated complex binds the promoter $P_{upt}$ with dissociation constant $K_{P}$ and thereby switches the transcription rate from its basal rate $\nu_m^0$ to its maximal rate $\nu_m$. The chemical reactions for transcriptional regulation are
\begin{eqnarray}
3\,a +  \mathrm{C}  & \stackrel{K_C}{\longleftrightarrow}  &  a_3\cdot \mathrm{C}\nonumber\\
 a_3\cdot \mathrm{C} + P_{upt} & \stackrel{K_P}{\longleftrightarrow}  &  a_3\cdot \mathrm{C}\cdot P_{upt} \nonumber\\
 P_{upt}  & \stackrel{\nu_m^0}{\longrightarrow}  &   P_{upt} + m\nonumber\\
 a_3\cdot \mathrm{C} \cdot P_{upt} & \stackrel{\nu_{\mathrm{m}}}{\longrightarrow}  & a_3\cdot  \mathrm{C}\cdot P_{upt}+ m \,.\nonumber
\end{eqnarray}
Here the concentration of AraC molecules [C] is a variable that changes little over time \cite{Johnson_JBacteriol_95} and is therefore assumed to be a constant parameter in our model. In steady state, the probability for finding the promoter $P_{upt}$ in a transcriptionally activated state is a Hill function of the internal arabinose concentration, $\left[ a_3\cdot C\cdot P_{upt} \right] = [a]^3/\left(K_C K_P/[C] + [a]^3 \right)$. We define the effective arabinose threshold for activation of $P_{upt}$ as $K_{upt}\equiv \left(K_C K_P/[C]\right)^{1/3}$.

{\it Translation \& turnover.} mRNA is translated into functional uptake protein at rate $\nu_p$ and gets degraded at rate $\lambda_m$. In contrast, the uptake proteins and arabinose are only diluted by cell growth at doubling rate $\gamma$:
\begin{eqnarray}
 m  & \stackrel{\nu_{p}}{\longrightarrow} & m +  \mathrm{Upt}\nonumber\\
 m &  \stackrel{\lambda_{m}}{\longrightarrow}  &\o\nonumber\\
  \mathrm{Upt} &  \stackrel{\gamma}{\longrightarrow}  & \o \nonumber \\
  a &  \stackrel{\gamma}{\longrightarrow}  & \o\,. \nonumber
\end{eqnarray}

{\it Delay time distribution.} Following Berg \cite{Berg_PNAS_78}, we derive an analytical approximation for the delay time distribution of our stochastic model. In the absence of arabinose, transcription of the gene for the uptake protein takes place at its basal rate $\nu_{m}^0$. Neglecting operator state fluctuations \cite{Kepler_BiophysJ_01}, the probability to observe $m$ transcription events up to time $t$ follows a Poisson distribution
\begin{equation}
\mathcal{P}(m|\nu_{m}^0 t) = \frac{(\nu_{m}^0 t)^m}{m!}e^{-\nu_{m}^0 t}\,, \nonumber
\end{equation}
with mean and variance $\nu_{m}^0 t$. In the limit of short mRNA lifetime $\lambda_m^{-1}$ compared to the protein lifetime $\lambda_p^{-1}$, one can assume instantaneous, geometrically distributed protein bursts from each mRNA molecule. This implies that the probability that $m$ mRNA molecules produce $n$ proteins follows a negative binomial distribution
\begin{equation}
\mathcal{NB}(n|m,b) = \left( \frac{1}{1+b}\right)^m\,\left( \frac{b}{1+b}\right)^n\,\left( \begin{array}{c} \mu+n-1\\n \end{array} \right)\,, \nonumber
\end{equation}
where the burst size $b\equiv\nu_p/\lambda_m$ is the average number of proteins produced from one mRNA molecule. Hence, the probability to produce $n$ proteins up to time $t$ is the weighed sum of negative binomials $P(n|\nu_m^0 t,b) \equiv \sum_m \mathcal{P}(m|\nu_m^0 t)\cdot \mathcal{NB} (n|m,b)$. Setting $t$ equal to the protein lifetime $\lambda_p^{-1}$ yields the steady state distribution of proteins, and for large $\mu\equiv \nu_m^0/\lambda_p$ we can  replace the Poisson distribution by a delta function located at $m=\mu$, leading to Eq.~\ref{EQNprotein_distribution} in the main text. Applying the transformation rule $Q(\tau_D)=\left| \frac{dn(\tau_D)}{d\tau_D}\right| P(n)$ yields the delay time distribution in Eq.~\ref{EQNdelay_time_distr} and the moments $\langle \tau_D \rangle$ and $\delta \tau_D^2  =  \langle \tau_D^2 \rangle - \langle \tau_D \rangle^2 $ are determined by the integrals
\begin{eqnarray}
\langle \tau_D \rangle & = & \int d\tau_D \tau_D Q(\tau_D) = \int dn \frac{\tau_0}{n} P(n)\;,\mathrm{and} \nonumber\\
\langle \tau_D^2 \rangle  & = & \int d\tau_D \tau_D^2 Q(\tau_D) = \int dn \frac{\tau_0^2}{n^2} P(n)\,.\nonumber
\end{eqnarray}
Here expansion of the integrands up to second order in $\delta n=n-\langle n\rangle$ brings us to Eqs.~\ref{EQNmean_var}.

{\it Parameter values.} The effective arabinose threshold $K_{upt}\approx50\,\mu$M  and the promoter binding constant $K_P = 10$\,nM are chosen similar to the parameters of $P_{BAD}$ \cite{Schleif_JMB_69,Timmes_JMB_04}. This choice determines the ratio $K_C/[\mathrm{C}] = K_{upt}^3/K_P $ (see above) and by choosing a typical value of $[\mathrm{C}] = 100$\,nM we obtain $K_C=10^6\,\mu $M$^3$. For the maximal promoter activity we set a typical value for the promoters in the {\it ara}-regulon, $\nu_{m}=5$ mRNA/min, which was derived from the mRNA steady state levels reported in \cite{Johnson_JBacteriol_95}. With a promoter fold-change of 150 similar to $P_E$ and $P_{FGH}$ \cite{Kolodrubetz_JMolBiol_81}, the basal transcription rate is expected to be on the order of $\nu_m^0\approx 0.03$\,mRNA/min . From our fits of Eq.~\ref{EQNdelay_time_distr} to the experimental delay time distributions we obtained an average value of $\mu=\nu_m^0/\lambda_p=3.8$. With our protein dilution rate of $\lambda_p=\gamma=\ln (2)/(50\,\mathrm{min})$ (from our measurement of the growth rate, see main text), this yields a basal expression rate of $\nu_m^0\approx0.05$\,mRNA/min - in good agreement with the biochemical constraints stated before. The mRNA degradation rate $\lambda_m$ is set according to a half-life of 2\,min \cite{Johnson_JBacteriol_95}, allowing us to adjust the translation rate $\nu_p$ to match a typical burst factor of $b=30$ \cite{Thattai_PNAS_01}. The $K_m$ for arabinose uptake is in wildtype cells about $50\,\mu$M \cite{Daruwalla_BiochemJ_81}, and the maximal uptake rate per uptake protein, $v_{max}$, can be estimated from bulk measurements in which the uptake rate per total cellular dry mass was determined \cite{Daruwalla_BiochemJ_81}. By assuming a dry mass of $3\cdot10^{-13}$\,g per cell \cite{Physiology_Cell} and about $10^3-10^4$ uptake proteins per cell \cite{Kehres_ProteinSci_92}, we end up with $v_{max}=200- 2000$ arabinose molecules/protein/min. From a Lineweaver-Burk fit to the data in Fig.~\ref{FIGfitted_parameters}~(b) we obtained $v_{max}\approx120$\,molecules/protein/min and $K_m=2.8$\,mM. While the value for $v_{max}$ is compatible with the biochemical constraints, our $K_m$ differs by two orders of magnitude from the previously reported value of $50\,\mu$M \cite{Daruwalla_BiochemJ_81}. For such a small Michaelis constant, all arabinose concentrations used in our experiments would saturate the uptake system completely and hence there should be no difference in timing of gene induction. However, the experimental conditions of Ref.~\cite{Daruwalla_BiochemJ_81} differ from ours; in particular, the proton gradient between periplasm and cytoplasm, which drives the arabinose/$H^+$ symport by AraE, is limited by oxygen availability \cite{Kashket_JBacteriol_81}. For the case of the lactose/$H^+$ symporter LacY, it has been shown that a reduced proton gradient leads to an increase of the apparent $K_m$ \cite{Kaback1997}. Hence, oxygen limitation in our micofluidic setup could explain the observed discrepancy.

{\it Stochastic simulations.} Although in the rate equations above only the equilibrium constants are depicted, we took for the dynamical simulations all association- and dissociation processes explicitly into account. As a conservative assumption, all association rates were chosen 10-fold smaller than the diffusion-limited on-rate of $2\, \mathrm{nM^{-1} min}^{-1}$ for a typical transcription factor in {\it E. coli} \cite{Bruinsma_PhysicaA_02} and the dissociation rates were adjusted according to the respective equilibrium constant. The trajectories in Fig.~\ref{FIGtimeseries_arabinose}~(b) and (c) correspond to single kinetic Monte-Carlo simulations \cite{Gillespie_JPhysChem_77} for 0.01\% external arabinose.
The protein and delay-time distributions in Fig.~\ref{FIGtimeseries_arabinose}~(a)~and~(d) (solid lines) were obtained from $5\cdot10^4$ independent simulation runs with the same parameters.

\normalsize

%%%%%%%%%%%%%%%%%%%%%%%%%%%%%%%%%%%%%%%%%%%%%

\section*{Acknowledgements} 
We are grateful to  R. Heermann for construction of the
plasmid. We thank T. Hwa for helpful discussions and M. Leisner for careful reading of the manuscript. 
This work was supported by the LMUinnovativ project
"Analysis and Modelling of Complex Systems". JM acknowledges
funding by the {\it Elitenetzwerk Bayern}. {\it Author
contributions:} JM carried out the experiments. GF performed the
simulations and analytical calculations. All authors designed the
research and wrote the paper.

\end{document}